\documentclass[sw]{AGUTeX}
\usepackage{fancyheadings}

\usepackage{listings}
\usepackage[dvips]{graphicx}

\def\rr#1{#1}

\def\ra#1{#1}

\authorrunninghead{SCHRIJVER EL AL}

\titlerunninghead{Space weather and electrical systems}

\authoraddr{C.J. Schrijver,
Lockheed Martin STAR Labs, 
3251 Hanover Street, Palo Alto, CA 94304, USA
(schrijver@lmsal.com)}

\authoraddr{R. Dobbins,
Risk Engineering Technical Strategies Team.
Zurich Services Corporation,
1400 American Lane,
Schaumburg, IL 60196, USA
(robert.dobbins@zurichna.com)}

\authoraddr{W. Murtagh,
Space Weather Prediction Center, NOAA,
325 Broadway/Room 2C112/W/NP9,
Boulder, CO 80305, USA
(william.murtagh@noaa.gov)}

\authoraddr{S.M. Petrinec,
Lockheed Martin STAR Labs, 
3251 Hanover Street, Palo Alto, CA 94304, USA
(petrinec@lmsal.com)}

\begin{document}
\pagestyle{fancy}
\cfoot{\thepage}
\centerline{\em Preprint for Space Weather Journal (2014)}


%
%

\title{Assessing the impact of space weather on the electric power grid based on insurance claims for industrial electrical equipment}

%

%
%



\authors{C. J. Schrijver\altaffilmark{1}, R. Dobbins\altaffilmark{2},
  W. Murtagh\altaffilmark{3}, S.M. Petrinec\altaffilmark{1}}

\altaffiltext{1}{Lockheed Martin Advanced Technology Center, 
Palo Alto, CA,  USA}
\altaffiltext{2}{Zurich Services Corporation, Schaumburg, IL,  USA}
\altaffiltext{3}{Space Weather Prediction Center, Boulder, CO,  USA}

\lefthead{Schrijver et al.}
\righthead{Space weather and insurance claims for electrical systems}

%
%


\begin{abstract}
  Geomagnetically induced currents are known to induce disturbances in
  the electric power grid. Here, we perform a statistical analysis of
  11,242 insurance claims from 2000 through 2010 for equipment losses
  and related business interruptions in North-American commercial
  organizations that are associated with damage to, or malfunction of,
  electrical and electronic equipment. We find that claims rates are
  elevated on days with elevated geomagnetic activity by approximately
  20\%\ for the top 5\%, and by about $10$\%\ for the top third of
  most active days ranked by daily maximum variability of the
  geomagnetic field. When focusing on the claims explicitly attributed
  to electrical surges (amounting to more than half the total sample),
  we find that the dependence of claims rates on geomagnetic activity
  mirrors that of major disturbances in the U.S.\ high-voltage
  electric power grid. The claims statistics thus reveal that
  large-scale geomagnetic variability couples into the low-voltage
  power distribution network and that related power-quality variations
  can cause malfunctions and failures in electrical and electronic
  devices that, in turn, lead to an estimated 500 claims per average year
  within North America. We discuss the possible magnitude of the full
  economic impact associated with quality variations in electrical
  power associated with space weather.
\end{abstract}

%
%

%

\begin{article}


%
%

\section{Introduction}
Large explosions that expel hot, magnetized gases on the Sun can, should
they eventually envelop Earth, effect severe disturbances in the
geomagnetic field. These, in turn, cause geomagnetically induced
currents (GICs) to run through the surface layers of the Earth and
through conducting infrastructures in and on these, including the
electrical power grids. \rr{The storm-related GICs run on a background of daily variations associated with solar (X)(E)UV irradiation that itself is variable through its dependence on both quiescent and flaring processes.}

The strongest GIC events are known to have
impacted the power grid on occasion \citep[see,
e.g.,][]{kappenmanetal1997,boteler+etal98,arslan+etal2002,kappenman2005,wik+etal2009}.
Among the best-known of such impacts is the 1989 Hydro-Qu{\'e}bec
blackout \citep[e.g.,][]{bolduc2002,beland+small2004}. Impacts are
likely strongest at mid to high geomagnetic latitudes, but
low-latitude regions also appear susceptible \citep{gaunt2013}.

The potential for severe impacts on the high-voltage power grid and
thereby on society that depends on it has been assessed in studies by
government, academic, and insurance industry working groups
\citep[e.g.,][]{severeswx2008,fema2010,kappenman2010,swximpactlloyds2011,jason2011}. How
costly such potential major grid failures would be remains to be
determined, but impacts of many billions of dollars have been
suggested \citep[e.g.,][]{severeswx2008,jason2011}.

Non-catastrophic GIC effects on the high-voltage electrical grid
percolate into financial consequences for the power market
\citep{forbesstcyr2004,forbesstcyr2008,forbesstcyr2010} leading to
price variations on the bulk electrical power market on the order of a
few percent \citep{forbesstcyr2004}.

\cite{2013JSWSC...3A..19S} quantified the susceptibility of the U.S.\
high-voltage power grid to severe, yet not extreme, space storms,
leading to power outages and power-quality variations related to
voltage sags and frequency changes. They find, ``with more than
3$\sigma$ significance, that approximately 4\%\ of the disturbances in
the US power grid reported to the US Department of Energy are
attributable to strong geomagnetic activity and its associated
geomagnetically induced currents.''

The effects of GICs on the high-voltage power grid can, in turn,
affect the low-voltage distribution networks and, in principle, might
impact electrical and electronic systems of users of those regional
and local networks. A first indication that this does indeed happen
was reported on in association with tests conducted by the Idaho
National Laboratory (INL) and the Defense Threat Reduction Agency
(DTRA). They reported \citep{INLGIC2013} that "INL and DTRA used the
lab's unique power grid and a pair of 138kV core form, 2 winding
substation transformers, which had been in-service at INL since the
1950s, to perform the first full-scale testing to replicate conditions
electric utilities could experience from geomagnetic disturbances." In
these experiments, the researchers could study how the artificial
GIC-like currents resulted in harmonics on the power lines that can
affect the power transmission and distribution equipment. These "tests
demonstrated that geomagnetic-induced harmonics are strong enough to
penetrate many power line filters and cause temporary resets to
computer power supplies and disruption to electronic equipment, such
as uninterruptible power supplies".

In parallel to that experiment, we collected information on insurance
claims submitted to Zurich North-America (NA) for damage to, or
outages of, electrical and electronic systems from all types of
industries for a comparison with geomagnetic variability.  Here, we
report on the results of a retrospective cohort exposure analysis of
the impact of geomagnetic variability on the frequency of insurance
claims.  \rr{In this analysis, we contrast insurance claims
  frequencies on ``high-exposure'' dates (i.e., dates of high
  geomagnetic activity) with a control sample of ``low-exposure''
  dates (i.e., dates with essentially quiescent space weather
  conditions), carefully matching each high-exposure date to a control
  sample nearby in time so that we may assume no systematic changes in
  conditions other than space weather occurred between the exposure
  dates and their controls (thus compensating for seasonal weather
  changes and other trends and cycles).}

For comparison purposes we repeat the analysis of the frequency of
disturbances in the high-voltage electrical power grid as performed by
\cite{2013JSWSC...3A..19S} for the same date range and with matching
criteria for threshold setting and for the selection of the control
samples. In Section~\ref{sec:data} we describe the insurance claim
data, the metric of geomagnetic variability used, and the
grid-disturbance information. The procedure to test for any impacts of
space weather on insurance claims and the high-voltage power grid is
presented and applied in Section~\ref{sec:testing}. We summarize our
conclusions in Section~\ref{sec:conclusions} where we also discuss the
challenges in translating the statistics on claims and disturbances
into an economic impact.

\begin{figure*}[b]
\noindent\includegraphics[width=15cm]{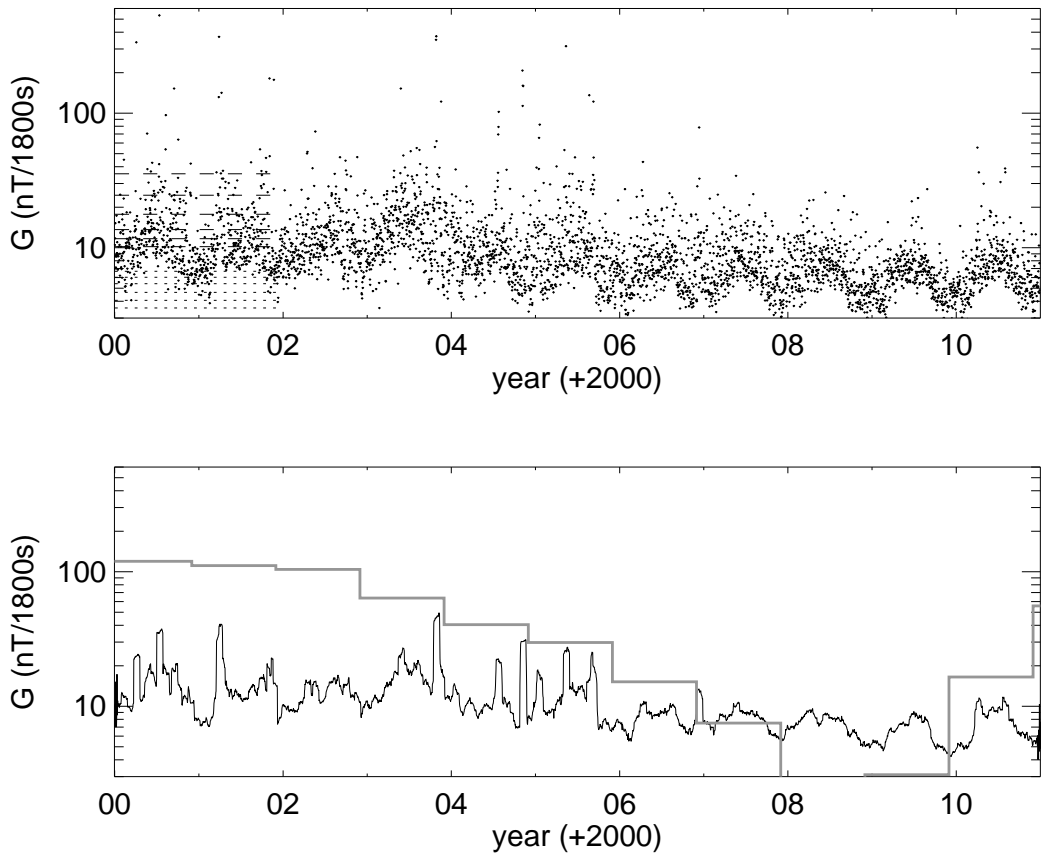} 
\caption{\em Daily values $G\equiv max(|dB/dt|)$ based on 30-min.\
  intervals (dots; nT/1800s) characterizing geomagnetic variability
  for the contiguous United States versus time (in years since
  2000). The 27-d running \rr{mean} is shown by the solid line. The
  levels for the 98, 95, 90, 82, 75, and 67 percentiles of the entire
  sample are shown by dashed lines (sorting downward from the top
  value of $G$) and dotted lines (sorting upward from the minimum
  value of the daily geomagnetic variability as expressed by $G\equiv
  max(|dB/dt|)$).  The grey histogram shows the annual \rr{mean}
  sunspot number.}
\label{fig:a}
\end{figure*}
\section{Data}\label{sec:data}
\subsection{Insurance claim data}\label{sec:claimsdata}
We compiled a list of all insurance claims filed by commercial
organizations to Zurich NA relating to costs incurred for electrical
and electronic systems for the 11-year interval from 2000/01/01
through 2010/12/31. Available for our study were the date of the event
to which the claim referred, the state or province within which the
event occurred, a brief description of the affected equipment, and a
top-level assessment of the probable cause. Information that might
lead to identification of the insured parties was not disclosed.

Zurich NA estimates that it has a market share of approximately 8\%\
in North America for policies covering commercially-used electrical
and electronic equipment and contingency business interruptions
related to their failure to function properly during the study
period. Using that information as a multiplier suggests that overall
some 12,800 claims are filed per average year related to
electrical/electronic equipment problems in North-American
businesses. The data available for this study cannot reveal impacts on
uninsured or self-insured organizations or impacts in events of which
the costs fall below the policy deductable.

The 11-year period under study has the same duration as that
characteristic of the solar magnetic activity cycle.  Fig.~\ref{fig:a}
shows that the start of this period coincides with the maximum in the
annual sunspot number for 2000, followed by a decline into an extended
minimum period in 2008 and 2009, ending with the rise of sunspot
number into the start of the next cycle.

The full sample of claims, regardless of attribution, for which an
electrical or electronic system was involved includes 11,242
entries. We refer to this complete set as set {\em A}.

\begin{figure}[t]
\noindent\includegraphics[width=8cm]{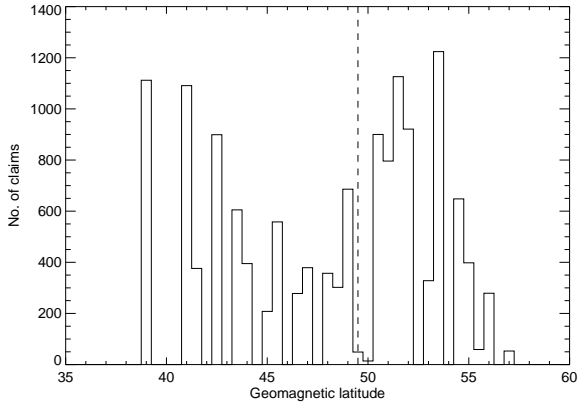} 
\caption{\em Number of insurance claims sorted by geomagnetic latitude
(using the \rr{central} geographical location of the state)  in 0.5$^\circ$ bins. The dashed
line at 49.5$^\circ$ is near the median geomagnetic latitude of the sample (at 49.3$^\circ$),
separating what this paper refers to as high-latitude from low-latitude states.}
\label{fig:c}
\end{figure}
Claims that were attributed to causes that were in all likelihood not
associated with space weather phenomena were deleted from set {\em A}
to form set {\em B} (with 8,151 entries remaining after review of the
Accident Narrative description of each line item). Such omitted claims
included attributions to water leaks and flooding, stolen or lost
equipment, vandalism or other intentional damage, vehicle damage or
vehicular accidents, animal intrusions (raccoons, squirrels, birds,
etc.), obvious mechanical damage, and obvious weather damage (ice
storm damage, hurricane/windstorm damage, etc.).  The probable causes
for the events making up set {\em B} were limited to the following
categories (sorted by the occurrence frequency, given in percent):
Misc: Electrical surge (59\%); Apparatus, Miscellaneous Electrical -
Breaking (30\%); Apparatus, Miscellaneous Electrical - Arcing (4.1\%);
Electronics - Breaking (1.6\%); Apparatus, Miscellaneous Electrical -
Overheating (1.4\%); Transformers - Arcing (0.9\%); Electronics -
Arcing (0.6\%); Transformers - Breaking (0.5\%); Generators - Breaking
(0.4\%); Apparatus, Electronics - Overheating (0.3\%); Generators -
Arcing (0.2\%); Generators - Overheating (0.2\%); and Transformers -
Overheating (0.1\%).

Fig.~\ref{fig:c} shows the number of claims received as a function of
the mean geomagnetic latitude for the state within which the claim was
recorded. Based on this histogram, we divided the claims into
categories of comparable size for high and low geomagnetic latitudes
along a separation at $49.{^\circ}5$ north geomagnetic latitude to
enable testing for a dependence on proximity to the auroral zones.
\rr{We note that we do not have access to information about the
  latitudinal distribution of insured assets, only on the claims
  received. Hence, we can only assess any dependence of insurance
  claims on latitude in a relative sense, comparing excess relative
  claims frequencies for claims above and below the median geomagnetic
  latitudes, as discussed in Sect.~\ref{sec:testing}.}

\begin{figure}[b]
\noindent\includegraphics[width=8cm]{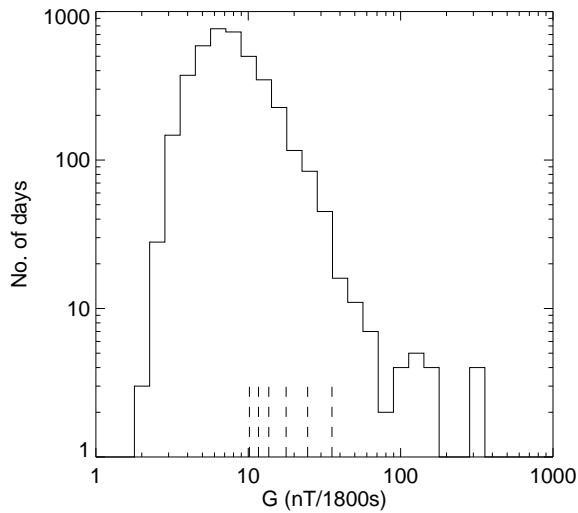} 
\caption{\em Histogram of the number of days between 2000/01/01 and
  2010/12/31 with values of $G\equiv max(|dB/dt|)$ in logarithmically
  spaced intervals as shown on the horizontal axis. The 98, 95, 90,
  82, 75, and 67 percentiles 
(ranking $G$ from low to high) are shown by dashed lines. }
\label{fig:b}
\end{figure}
\subsection{Geomagnetic data}
\ra{Geomagnetically-induced currents are driven by changes in
  the geomagnetic field. These changes are caused by the interaction
  of the variable, magnetized solar wind with the geomagnetic field and
  by the insolation of Earth's atmosphere that varies globally with
  solar activity and locally owing to the Earth's daily rotation and
  annual revolution in its orbit around the Sun. A variety of
  geomagnetic activity indices is available to characterize
  geomagnetic field variability \cite[e.g.,][]{handbook}. These
  indices are sensitive to different aspects of the variable
  geomagnetic-ionospheric current systems as they may differentially
  filter or weight storm-time variations (Dst), disturbance-daily
  variations (Ds), or solar quiet daily variations (known as the Sq
  field), and may weight differentially by (geomagnetic)
  latitude. Here, we are interested not in any particular driver of
  changes in the geomagnetic field but rather need a metric of the
  rate of change in the strength of the surface magnetic field as that
  is the primary driver of geomagnetically-induced currents.}

To quantify \ra{the variability in the
  geomagnetic field} we use the same metric as
\cite{2013JSWSC...3A..19S} based on the minute-by-minute geomagnetic
field measurements from the Boulder (BOU) and Fredericksburg (FRD)
stations (available via http://ottawa.intermagnet.org): we use these
measurements to compute the daily maximum value, $G$, of $|dB/dt|$
over 30-min.\ intervals, using the \rr{mean} value for the two
stations. \rr{We selected this metric recognizing a need to use a more
  regional metric than the often-used global metrics, but also
  recognizing that the available geomagnetic and insurance claims data
  have poor geographical resolution so that a focus on a metric
  responsive to relatively low-order geomagnetic variability was
  appropriate. We chose a time base short enough to be sensitive to
  rapid changes in the geomagnetic field, but long enough that it is
  also sensitive to sustained changes over the course of over some
  tens of minutes. For the purpose of this study, we chose to use a single metric of
  geomagnetic variability, but with the conclusion of our pilot study
  revealing a dependence of damage to electrical and electronic
  equipment on space weather conditions, a multi-parameter follow up
  study is clearly warranted, ideally also with more information on
  insurance claims, than could be achieved with what we have access to
  for this exploratory study.}

The
BOU and FRD stations are located along the central latitudinal axis of
the U.S.. The averaging of their measurements somewhat emphasizes
the eastern U.S.\ as do the grid and population that uses
that. Because the insurance claims use dates based on local time we
compute the daily $G$ values based on date boundaries of U.S.\ central
time. Fig.~\ref{fig:b} shows the distribution of values of $G$, while
also showing the levels of the percentiles for the rank-sorted value
of $G$ used as threshold values for a series of sub-samples in the
following sections.

\begin{figure}[b]
\noindent\includegraphics[width=8cm]{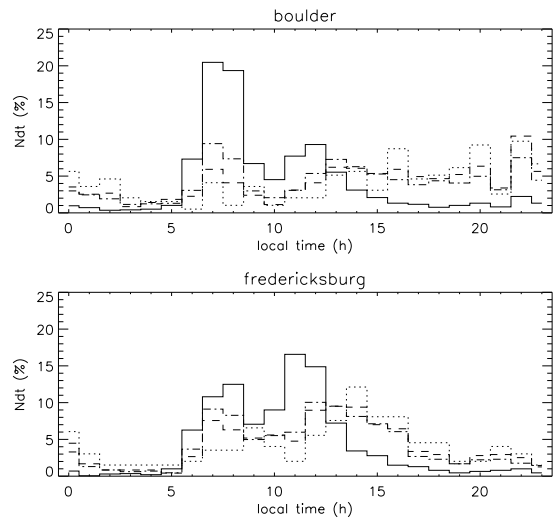} 
\caption{\em \rr{Normalized histograms of the local times for which
    the values of $G\equiv max(|dB/dt|)$ reach their daily maximum
    (top: Boulder; bottom: Fredericksburg). The solid histogram shows
    the distribution for daily peaks for all dates with $G$ values in
    the lower half of the distribution, i.e., for generally quiescent
    conditions. The dotted, dashed, and dashed-dotted histograms show
    the distributions for dates with high $G$ values, for thresholds
    set at the 95, 82, and 67 percentiles of the set of values for
    $G$, respectively.}}
\label{fig:N1}
\end{figure}
\rr{Figure~\ref{fig:N1} shows the local times at which the maximum
  variations in the geomagnetic field occur during 30-min.\
  intervals. The most pronounced peak in the distribution for
  geomagnetically quiet days (solid histogram) occurs around $7-8$
  o'clock local time, i.e., a few hours after sunrise, and a second
  peak occurs around local noon. The histograms for the subsets of
  geomagnetically active days for which $G$ values exceed thresholds
  set at 67, 82, and 95 percentiles of the sample are much broader,
  even more so for the Boulder station than for the Fredericksburg
  station.  From the perspective of the present study, it is important
  to note that the majority of the peak times for our metric of
  geomagnetic variability occurs within the economically most active
  window from 7 to 18 hours local time; for example, at the
  82-percentile of geomagnetic variability in $G$, 54\%\ and 77\%\ of
  the peak variability occur in that time span for Boulder and
  Fredericksburg, respectively.}

\ra{From a general physics perspective, we note that periods of
  markedly enhanced geomagnetic activity ride on top of a daily
  background variation of the ionospheric current systes (largely
    associated with the ``solar quiet'' modulations, referred to as
    the Sq field) that is induced to a large extent by solar
    irradiation of the atmosphere of the rotating Earth, including the
    variable coronal components associated with active-region gradual
    evolution and impulsive solar flaring. We do not attempt to
    separate the impacts of these drivers in this study, both because
    we do not have information on the local times for which the
    problems occurred that lead to the insurance claims, and because
    the power grid is sensitive to the total variability in the
    geomagnetic field regardless of cause.}

The daily $G$ values are shown versus time in Fig.~\ref{fig:a}, along
with a 27-d running \rr{mean} and (as a grey histogram) the yearly
sunspot number.  As expected, the $G$ value shows strong upward
excursions particularly during the sunspot maximum. 
Note the annual modulation in $G$ with generally
lower values in the northern-hemispheric winter months than in the
summer \rr{months.}

\subsection{Power-grid disturbances}
In parallel to the analysis of the insurance claims statistics, we
also analyze the frequencies of disturbances in the U.S.\ high-voltage
power grid.  \cite{2013JSWSC...3A..19S} compiled a list of ``system
disturbances'' published by the North American Electric Reliability
Corporation (NERC: available since 1992) and by the Office of
Electricity Delivery and Energy Reliability of the Department of
Energy (DOE; available since 2000). This information is compiled by
NERC for a region with over 300 million electric power customers
throughout the U.S.A.\ and in Ontario and New Brunswick in Canada,
connected by more than 340,000\,km of high-voltage transmission lines
delivering power generated in some 18,000 power plants within the
U.S.\ \citep{jason2011}.  The reported disturbances include, among
others, ``electric service interruptions, voltage reductions, acts of
sabotage, unusual occurrences that can affect the reliability of the
bulk electric systems, and fuel problems.''  We use the complete set
of disturbances reported from 2000/01/01 through 2010/12/31 regardless
of attributed cause. We refer to \cite{2013JSWSC...3A..19S} for more
details.

\begin{figure*}
\noindent\includegraphics[width=8cm]{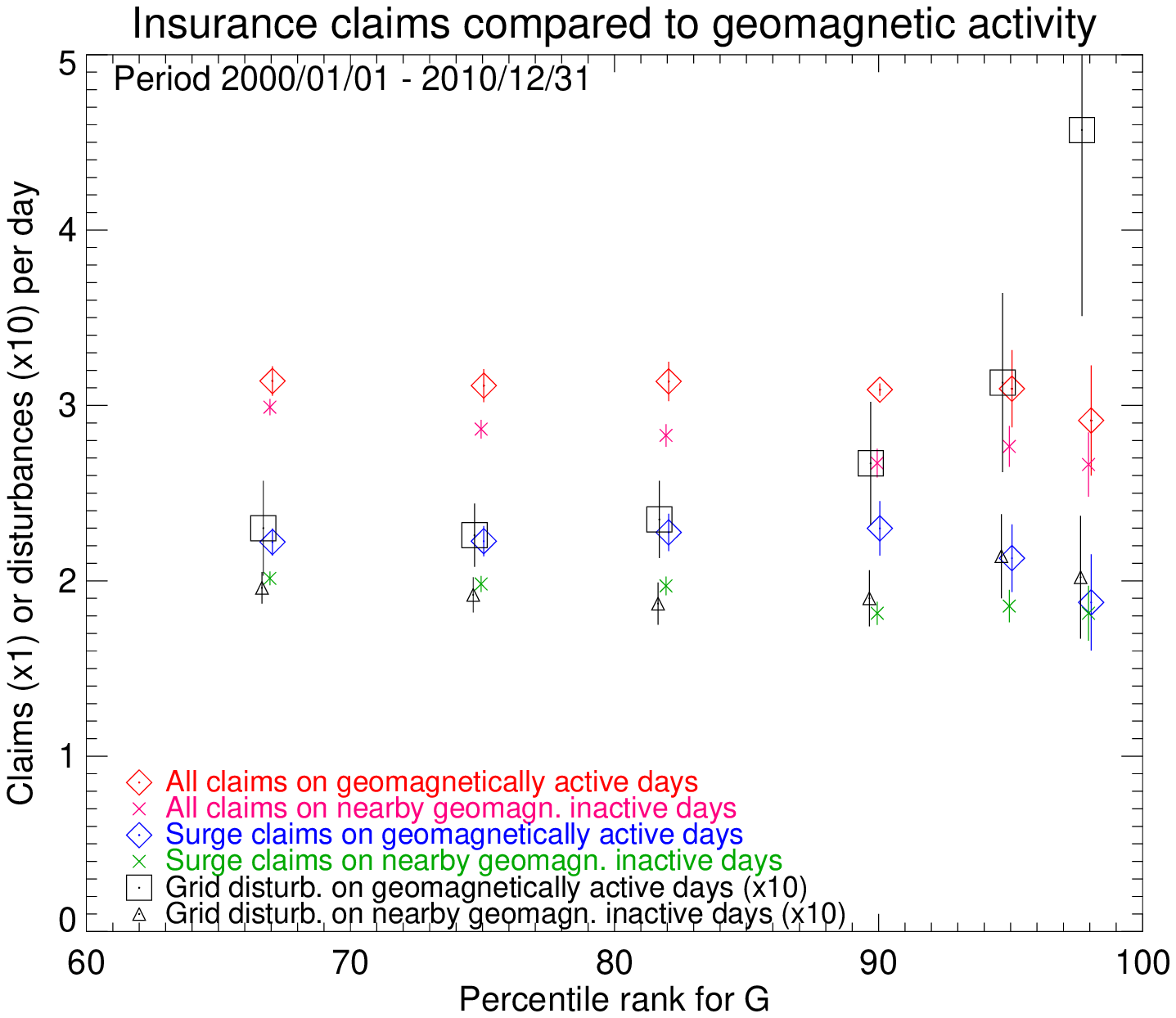}\includegraphics[width=8cm]{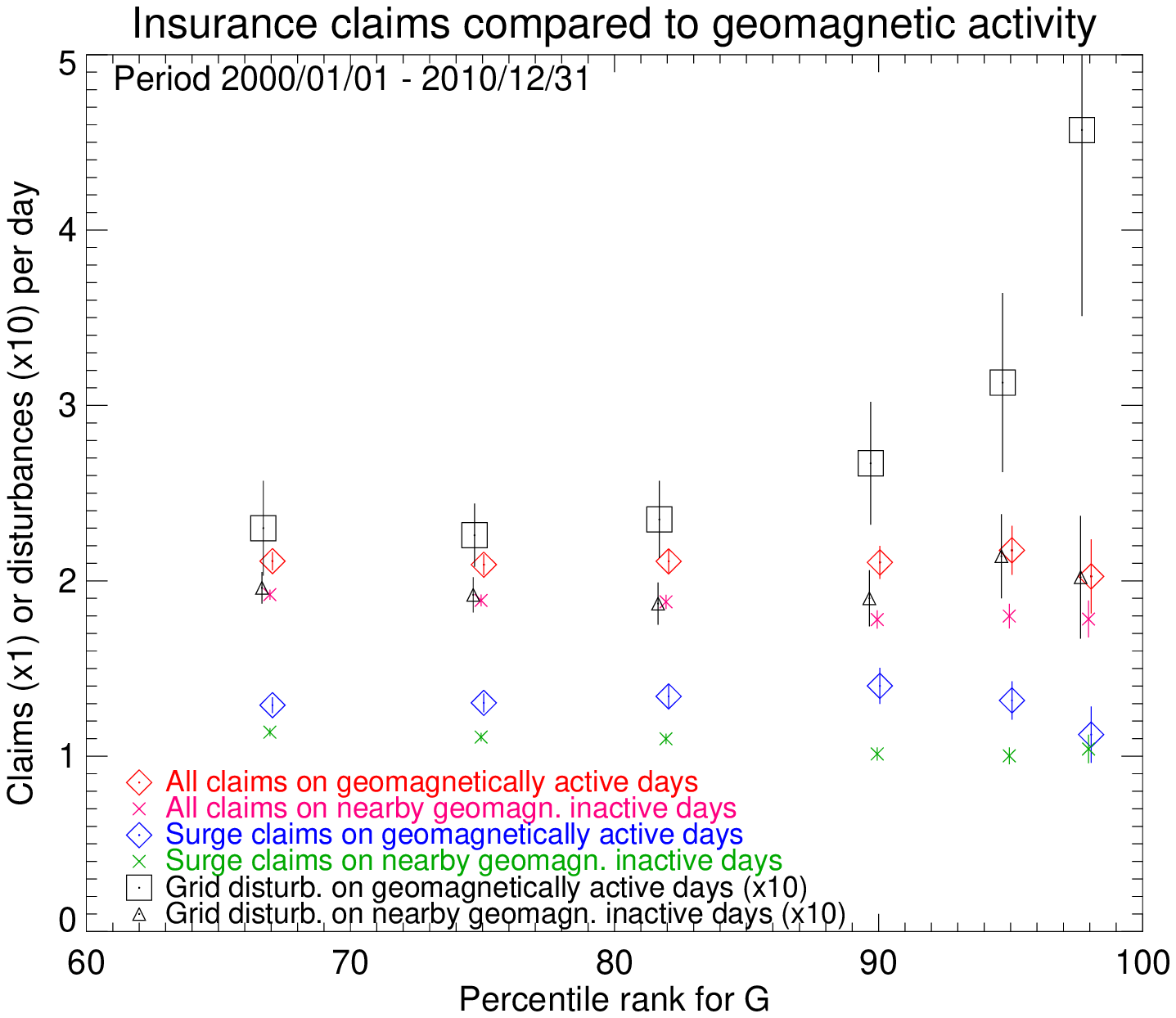} 
\caption{\em Claims per day for the full sample of insurance claims
  (set {\em A} left) and the sample from which claims likely unrelated
  to any space weather influence have been removed (set {\em B},
  right). Each panel shows \rr{mean} incident claim frequencies $n_i\pm
  \sigma_c$ (diamonds) for the most geomagnetically active dates,
  specifically for the 98, 95, 90, 82, 75, and 67 percentiles of the
  distribution of daily values of $G\equiv max(|dB/dt|)$ sorted from
  low to high (shown with slight horizontal offsets to avoid overlap
  in the symbols and \rr{bars showing the standard deviations for the mean values}). The asterisks show the
  associated claim frequencies $n_c \pm \sigma_c$, for the control
  samples. The panels also show the frequencies of reported high-voltage
  power-grid disturbances (diamonds and triangles for geomagnetically
  active dates and for control dates, respectively), multiplied by 10
  for easier comparison, using the same exposure-control sampling and
  applied to the same date range as that used for the insurance
  claims.}
\label{fig:d}
\end{figure*}

\section{Testing for the impact of space weather}\label{sec:testing}
In order to quantify effects of geomagnetic variability on the frequency
of insurance claims filed for electrical and electronic equipment we
need to carefully control for a multitude of variables that include
trends in solar activity, the structure and operation of the power
grid \ra{(including,
  for example, scheduled maintenance and inspection), various societal and technological factors changing over the
years, as well as the costs and procedures related to the insurance
industry, and, of course, weather and seasonal trends related to
  the insolation angle and the varying tilt of the Earth's magnetic field
  relative to the incoming solar wind throughout the year}. 

\nocite{wacholderetal1992} \nocite{schulz+grimes2002}
\nocite{grimes+schulz2005} \nocite{2013JSWSC...3A..19S}

\rr{There are many parameters that may influence the ionospheric
  current systems, the quality and continuity of electrical power, and
  the malfunctioning of equipment running on electrical power. We may
  not presume that we could identify and obtain all such parameters,
  or that all power grid segments and all equipment would respond
  similarly to changes in these parameters. We therefore do not
  attempt a multi-parameter correlation study, but instead apply a
  retrospective cohort exposure study with tightly matched controls
  very similar to that applied by Schrijver and Mitchell (2013).}

\rr{This type of exposure study is based on pairing dates of exposure,
  i.e., of elevated geomagnetic activity, with control dates of low
  geomagnetic activity shortly before or after each of the dates of
  exposure, selected from within a fairly narrow window in time during
  which we expect no substantial systematic variation in ionospheric
  conditions, weather, the operations of the grid, or the equipment
  powered by the grid.  Our results are based on a comparison of
  claims counts on exposure dates relative to claims counts on
  matching sets of nearby control dates. This minimizes the impacts of
  trends (including ``confounders'') in any of the potential factors
  that affect the claims statistics or geomagnetic variability,
  including the daily variations in quiet-Sun irradiance and the
  seasonal variations as Earth orbits the Sun, the solar cycle, and
  the structure and operation of the electrical power network.  This
  is a standard method as used in, e.g., epidemiology. We refer to
  Wacholder et al.\ (1992, and references therein) for a discussion on
  this method particularly regarding ensuring of time comparability of
  the "exposed" and control samples, to Schulz and Grimes (2002) for a
  discussion on the comparison of cohort studies as applied here
  versus case-control studies, and to Grimes and Schulz (2005) for a
  discussion of selection biases in samples and their controls
  (specifically their example on pp. 1429-1430).}

We define a series of values of geomagnetic variability in order to
form sets of dates including different ranges of exposure, i.e., of
geomagnetic variability, \rr{so that each high exposure date is
  matched by representative low exposure dates as controls}. We create
exposure sets by selecting a series of threshold levels corresponding
to percentages of all dates with the most intense geomagnetic activity
as measured by the metric $G$. Specifically, we determined the values
of $G$ for which geomagnetic activity, sorted from least active
upward, includes 67\%, 75\%, 82\%, 90\%, 95\%, and 98\%\ of all dates
in our study period. For each threshold value we selected the dates
with $G$ exceeding that threshold (with possible further selection
criteria as described below). For each percentile set we compute the
\rr{mean} daily rate of incident claims, $n_i$, as well as \rr{the
  standard deviation on the mean}, $\sigma_i$, as determined from the
events in the day-by-day claims list.

In order to form tightly matched control samples for low ``exposure'',
we then select 3 dates within a 27-d period centered on each of the
selected high-activity days. The 27-d period, also known as the
Bartels period, is that characteristic of a full rotation of the solar
large-scale field as viewed from the orbiting Earth; $G$ values within
that period sample geomagnetic variability as induced during one full
solar rotation. This window for control sample selection is tighter
than that used by \cite{2013JSWSC...3A..19S} who used 100-day windows
centered on dates with reported grid disturbances. For the present
study we selected a narrower window to put even stronger limits on the
potential effects of any possible long-term trends in factors that
might influence claims statistics or geomagnetic variability.  We note
that there is no substantive change in our main conclusions for
control windows at least up to 100 days in duration.

The three dates selected from within this 27-d interval are
those with the lowest value of $G$ smoothed with a 3-day running
mean. We determine the \rr{mean} claim rate, $n_c$, for this control set
and the associated standard deviation in the mean, $\sigma_c$.

Fig.~\ref{fig:d} shows the resulting daily frequency of claims and the
standard deviations in the mean, $n_i\pm \sigma_i$, for the selected
percentiles, both for the full sample {\em A} (left panel) and for
sample {\em B} (right panel) from which claims were omitted that were
attributed to causes not likely associated directly or indirectly with
geomagnetic activity. For all percentile sets we see that the claim
frequencies $n_i$ on geomagnetically active days exceed the
frequencies $n_c$ for the control dates.

\begin{figure*}
\centerline{\noindent\includegraphics[width=12cm]{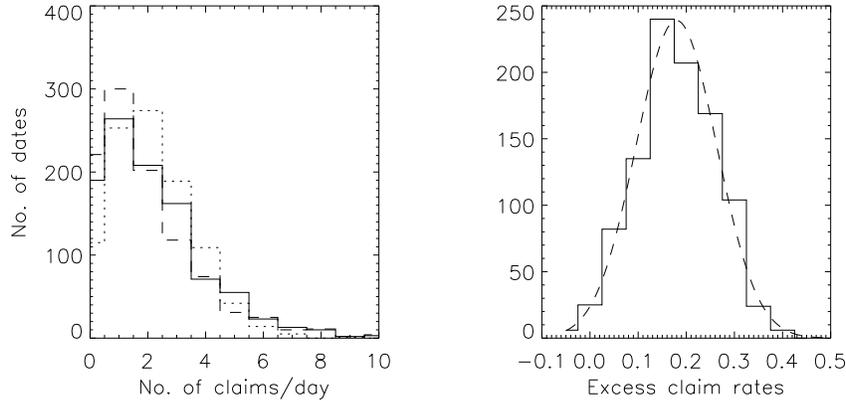}}
\caption{\em \rr{(left) Distribution of the number of claims per geomagnetically active day for set {\em B} for the top 25\%\ of $G$ values (solid) compared to that for the distribution of control dates (divided by 3 to yield the same total number of dates; dashed). For comparison, the expected histogram for a random Poisson distribution with the same mean as that for the geomagnetically active days is also shown (dotted). (right) Distribution (solid) of excess daily claim frequencies during geomagnetically active days (defined as in the left panel) over those on control dates determined by repeated random sampling from the observations (known as the bootstrap method), compared to a Gaussian distribution (dashed) with the same mean and standard deviation.} }
\label{fig:stats}
\end{figure*}
\rr{The frequency distributions of insurance claims are not Poisson
  distributions, as can be seen in the example in Fig.~\ref{fig:stats}
  (left panel): compared to a Poisson distribution of the same mean,
  the claims distributions on geomagnetically active dates, $N_{\rm
    B,a,75}$ and for control days, $N_{\rm B,c,75}$, are skewed to
  have a peak frequency at lower numbers and a raised tail at higher
  numbers; a Kolmogorov-Smirnov (KS) test suggests that the probability
  that $N_{\rm B,c,75}$ is consistent with a Poisson distribution with
  the same mean is 0.01 for this example.} \ra{The elevated tail of
  the distribution relative to a Poisson distribution suggests some
  correlation between claims events, which is of interest from an
  actuarial perspective as it suggests a nonlinear response of the
  power system to space weather that we cannot investigate further
  here owing to the signal to noise ratio of the results given our sample.}

\rr{For the case shown in
  Fig.~\ref{fig:stats} for the 25\%\ most geomagnetically active dates
  in set {\em B}, a KS test shows that the probability that $N_{\rm
    B,a,75}$ and $N_{\rm B,c,75}$ are drawn from the same parent
  distribution is of order $10^{-14}$, i.e. extremely unlikely.}

\begin{figure}[t]
\noindent\includegraphics[width=8cm]{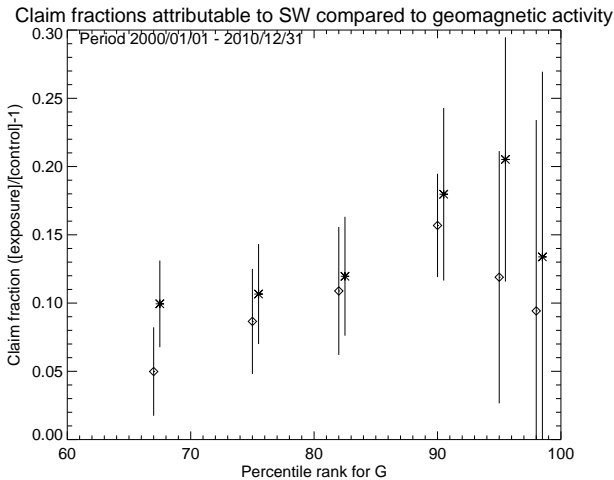} 
\caption{\em Relative excess claim frequencies \rr{statistically associated with} geomagnetic activity
  (difference between claim frequencies on geomagnetically active dates and
  the frequencies on control dates as shown in Fig.~\ref{fig:d}, i.e.,
  $(n_i-n_c)/n_c$) for the full sample (A; diamonds) and for the
  sample (B; asterisks) from which claims were removed attributable to
  apparently non-space-weather related causes.}
\label{fig:g}
\end{figure}

\begin{figure}[t]
\noindent\includegraphics[width=8cm]{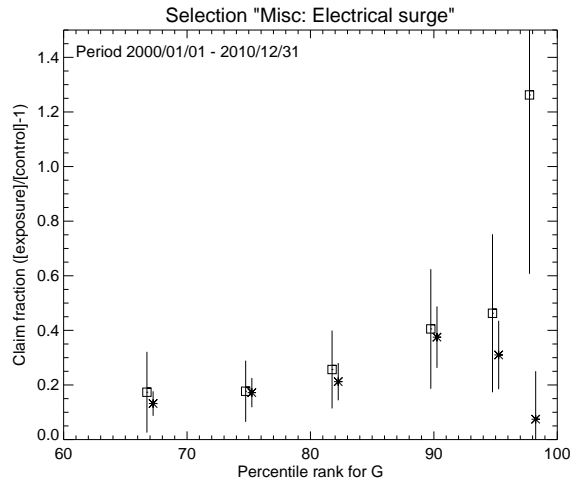} 
\caption{\em Same as Fig.~\ref{fig:g} but for sample {\em B} limited to those
claims attributed to ``Misc.: Electrical surge''  (asterisks) (for
57\%\ of the cases in that sample),
compared to the fraction of high-voltage power-grid disturbances
\rr{statistically associated with} geomagnetic activity (squares). }
\label{fig:g1}
\end{figure}

\begin{figure*}
\noindent\includegraphics[width=8cm]{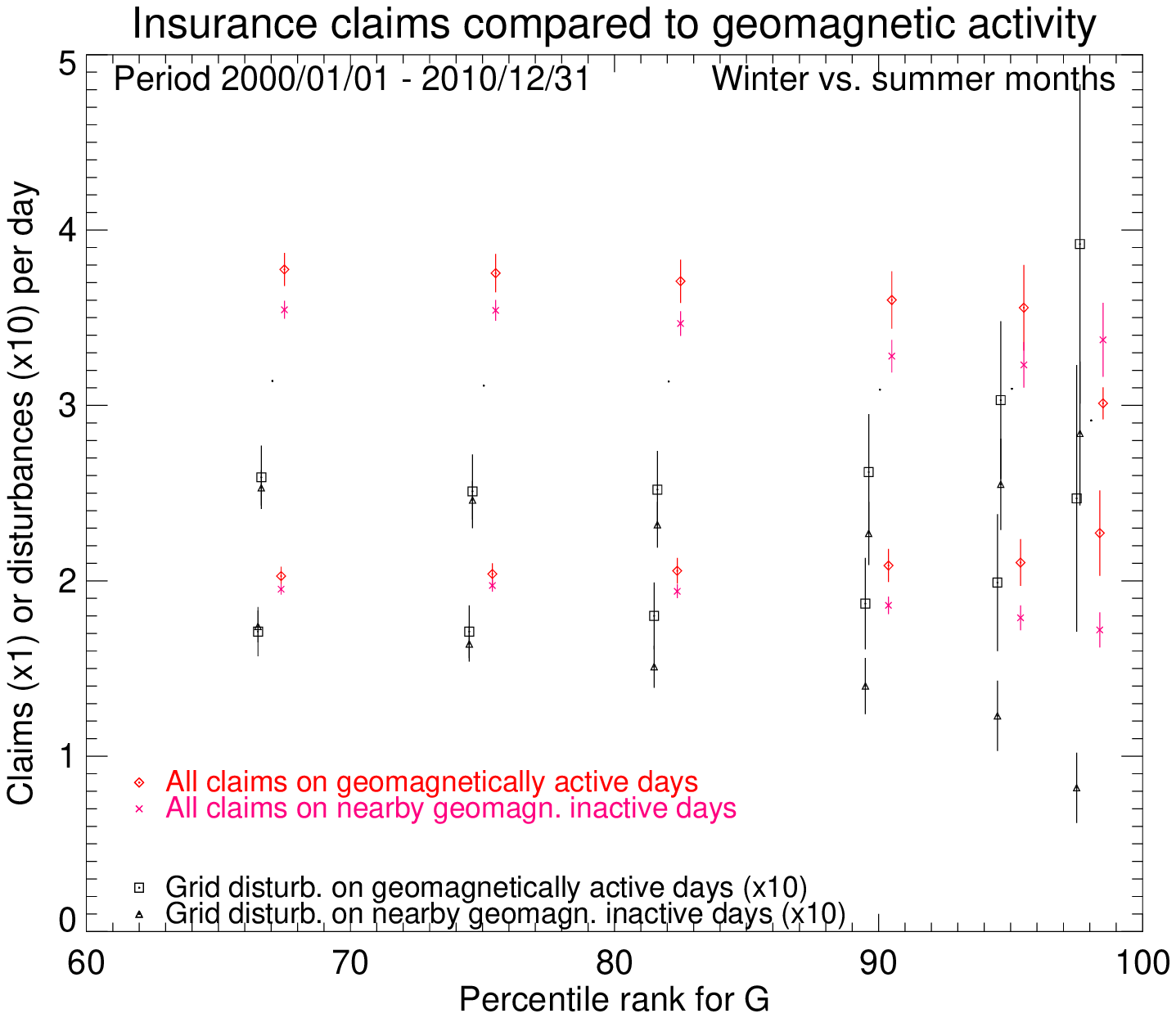}\includegraphics[width=8cm]{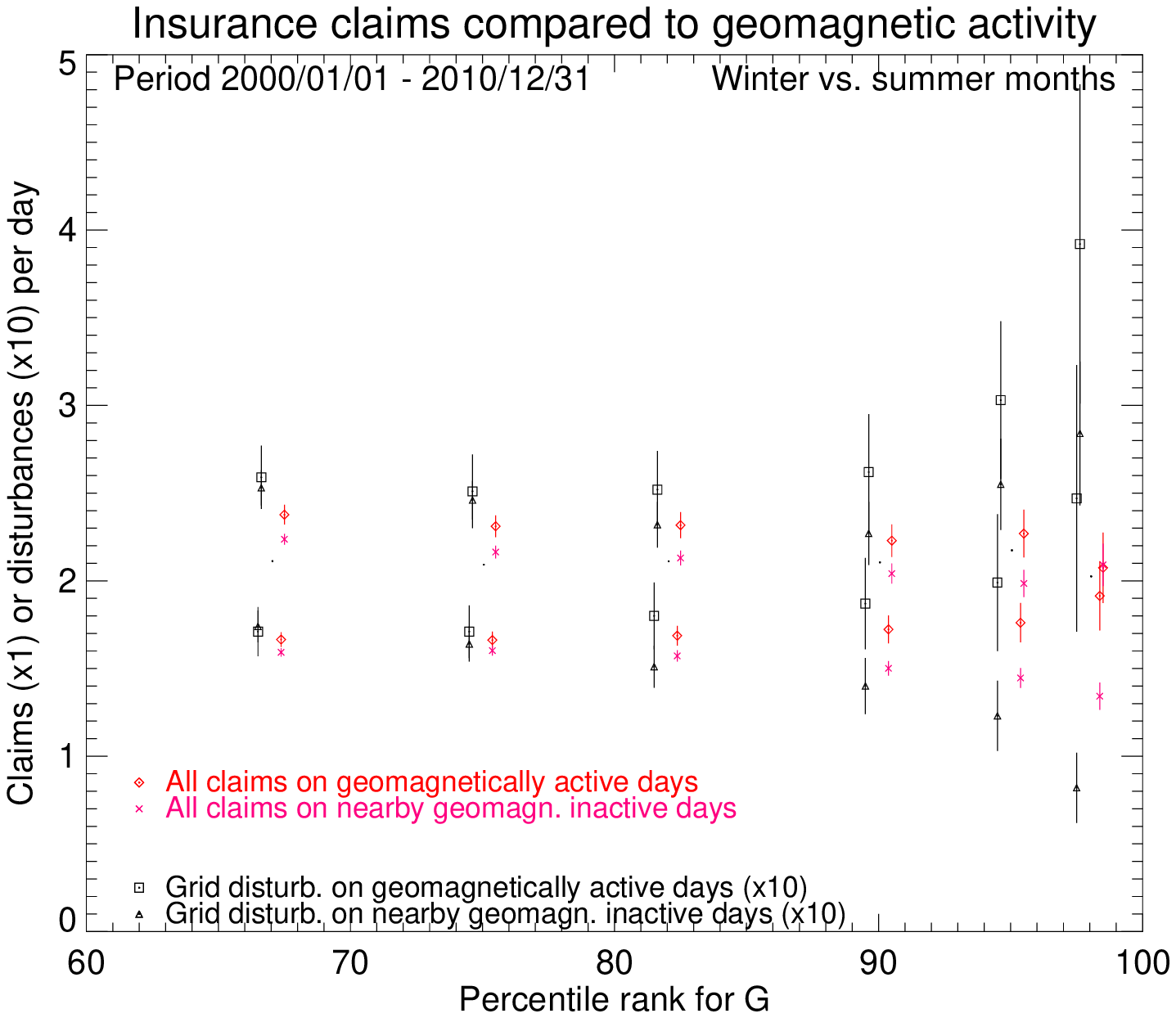} 
\caption{\em As Fig.~\ref{fig:d} but separating the winter half year
  (October through March) from the summer half year (April through
  September), for the full sample of insurance claims (set {\em A},
  left) and the sample from which claims likely unrelated to any space
  weather influence have been removed (set {\em B}, right). Values for
  the summer months are shown offset slightly
  towards the left of the percentiles tested (98, 95, 90, 82, 75, and
  67) while values for the winter months are offset to the
  right. Values for the winter season are systematically higher than
  those for summer months.}
\label{fig:e}
\end{figure*}

\begin{figure}
\noindent\includegraphics[width=8cm]{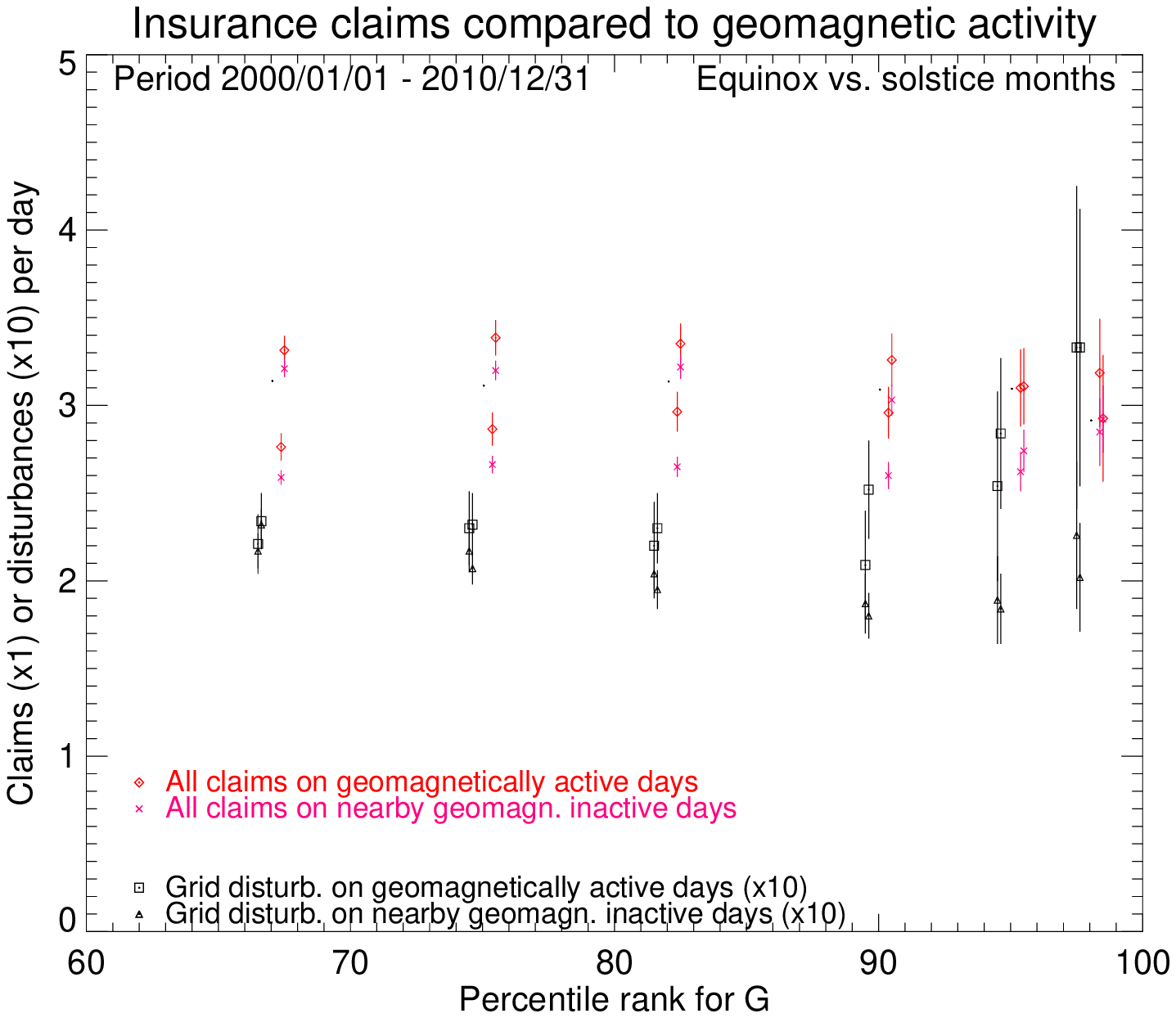}
\caption{\em \rr{As Fig.~\ref{fig:e} but separating the months around
    the equinoxes (February-April and August-October) from the
    complementing months around the solstices, for the full sample of
    insurance claims (set {\em A}). Values for the equinox periods are
    shown offset slightly towards the left of the percentiles tested
    (98, 95, 90, 82, 75, and 67) while values for the solstice months
    are offset to the right. Mean claims frequencies for the solstice
    periods are systematically higher than those for equinox periods,
    but the frequencies for high-$G$ days in excess of the control sample
    frequencies is slightly larger around the equinoxes than around the
    solstices.}}
\label{fig:e1}
\end{figure}

\begin{figure*}[b]
\noindent\includegraphics[width=8cm]{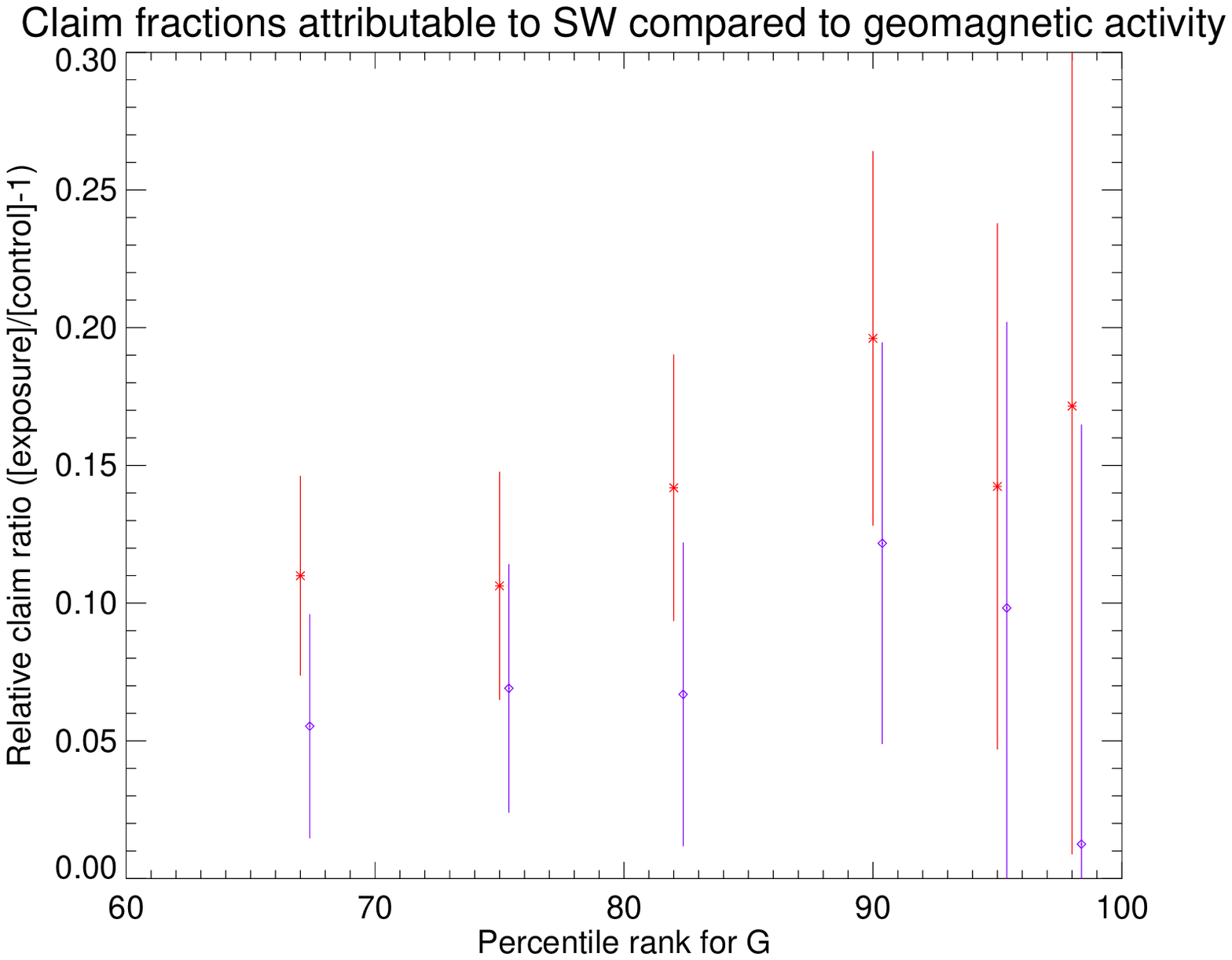}\includegraphics[width=8cm]{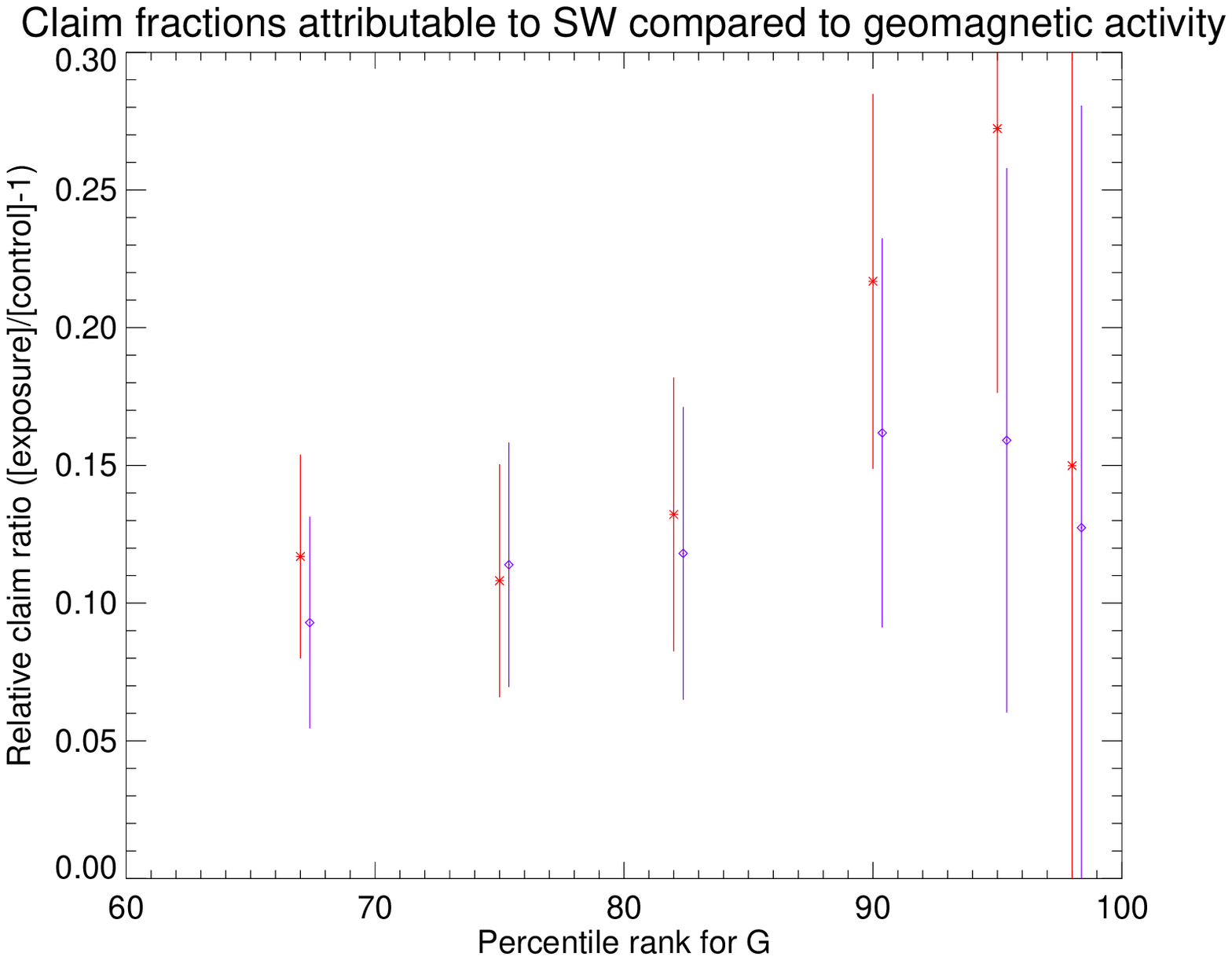} 
\caption{\em Relative excess claim frequencies $(n_i-n_c)/n_i$ on
  geomagnetically active dates relative to those on control dates for
  geomagnetic latitudes below $49.{^\circ}5$\,N (asterisks, red)
  compared to those for higher latitudes (diamonds, purple; offset
  slightly to the right) for the percentiles tested (98, 95, 90, 82,
  75, and 67). The lefthand panel shows the results for the full
  sample (A), and the righthand panel shows these for sample B from
  which apparently non-space-weather related events were removed (see
  Section~2.1).}
\label{fig:f}
\end{figure*}

\rr{The numbers that we are ultimately interested in are the excess
  frequencies of claims on geomagnetically active dates over those on the
  control dates, and their uncertainty. For the above data set, we
  find and excess daily claims rate of $(n_{\rm B,i}-n_{\rm B,c})\pm
  \sigma_{\rm B}=0.20\pm 0.08$. The uncertainty $\sigma_{\rm B}$ is in
  this case determined by repeated random sampling of the claims
  sample for exposure and control dates, and subsequently determining
  the standard deviation in a large sample of resulting excess frequencies
  (using the so-called bootstrap method). The distribution of excess
  frequencies (shown in the righthand panel of Fig.~\ref{fig:stats}) is
  essentially Gaussian, so that the metric of the standard deviation
  gives a useful value to specify the uncertainty. We note that the
  value of $\sigma_{\rm B}$ is comparable to the value $\sigma_{a,c}=
  (\sigma^2_{\rm a}+\sigma^2_{\rm b})^{1/2}$ derived by combining the
  standard deviations for the numbers of claims per day for
  geomagnetically active dates and the control dates, which in this
  case equals $\sigma_{a,c}=0.07$. Thus, despite the skewness of the
  claim count distributions relative to a Poisson distribution as
  shown in the example in the left panel of Fig.~\ref{fig:stats}, the
  effect of that on the uncertainty in the excess claims rate is
  relatively small. For this reason, we show the standard deviations
  on the mean frequencies in Figs.~\ref{fig:d}-\ref{fig:f} as a useful visual
  indicator of the significance of the differences in mean frequencies.}

Fig.~\ref{fig:g} shows the relative excess claims frequencies, i.e., the
relative differences $r_e=(n_i-n_c)/n_c$ between the claim frequencies on
geomagnetically active dates and those on the control dates, thus
quantifying the claim fraction \rr{statistically associated with}
elevated geomagnetic activity.  \rr{The uncertainties shown are
  computed as
  $\sigma_e=(\sigma_i^2/n_i^2+\sigma_c^2/n_c^2)^{1/2}\,r_e$, i.e.,
  using the approximation of normally distributed uncertainties,
  warranted by the arguments above.} We note that the relative
rate of claims \rr{statistically associated with} space weather is
slightly higher for sample {\em B} than for the full set {\em A}
consistent with the hypothesis that the claims omitted from sample
{\em A} to form sample {\em B} were indeed preferentially unaffected
by geomagnetic activity.  Most importantly, we note that the rate of
claims \rr{statistically associated with} geomagnetic activity
increases with the magnitude of that activity.

\rr{About 59\%\ of the claims in sample {\em
  B} attribute the case of the problem to ``Misc.: Electrical surge'', so that we can be certain that some variation in the quality or continuity of 
electrical power was involved. Fig.~\ref{fig:g1} shows the relative excess claims rate $(n_i-n_c)/n_c$
as function of threshold for geomagnetic activity.
We compare these results} with the same metric, based on identical
selection procedures, for the frequency of disturbances in the
high-voltage power grid (squares). We note that these two metrics, one
for interference with commercial electrical/electronic equipment and
one for high-voltage power, agree within the uncertainties, with the
possible exception of the infrequent highest geomagnetic activity (98
percentile) although there the statistical uncertainties on the mean frequencies
are so large that the difference is less than 2 standard
deviations \rr{in the mean values}.

\begin{table}
\caption{Probability ($p$) values based on a Kolmogorov-Smirnov test that the observed sets of claims numbers on geomagnetically active dates and on control dates are drawn from the same parent distribution, for date sets with the geomagnetic activity metric $G$ exceeding the percentile threshold in the distribution of values.}\label{tab:pvals}
\begin{tabular}{c | c  c | c c}
\hline
Percentile & \multicolumn{2}{|c}{All claims} & \multicolumn{2}{|c}{Attr. to electr. surges} \\
       & set {\em A} & set {\em B} &     set {\em A} & set {\em B} \\
\hline
67 & 2.$\times 10^{-10}$ & 2.$\times 10^{-19}$ & 1.$\times 10^{-27}$ & 0 \\
75 & 3.$\times 10^{-7}$ & 4.$\times 10^{-14}$ & 8.$\times 10^{-20}$ & 4.$\times 10^{-35}$ \\
82 & 0.0004 & 2.$\times 10^{-7}$ & 1.$\times 10^{-13}$ & 6.$\times 10^{-24}$ \\
90 & 0.010 & 0.0002 & 1.$\times 10^{-7}$ & 8.$\times 10^{-13}$ \\
95 & 0.05 & 0.013 & 0.0001 & 2.$\times 10^{-7}$ \\
98 & 0.33 & 0.06 & 0.003 & 0.0001\\ 
\hline
\end{tabular}
\end{table}
\rr{To quantify the significance of the excess claims frequencies on geomagnetically active days we perform a non-parametric Kolmogorov-Smirnov (KS) test of the null hypothesis that the claims events on active and on control days could be drawn from the same parent sample. The resulting $p$ values from the KS test, summarized in Table~\ref{tab:pvals}, show that it is extremely unlikely that our conclusion that geomagnetic activity has an impact on insurance claims could be based on chance, except for the highest percentiles in which the small sample sizes result in larger uncertainties. We note that the $p$ values tend to decrease when we eliminate claims most likely unaffected by space weather (contrasting set {\em A} with {\em B}) and when we limit either set to events attributed to electrical surges: biasing the sample tested towards issues more likely associated with power-grid variability increases the significance of our findings that there is an impact of space weather.}

Fig.~\ref{fig:e}  shows insurance claims differentiated by season:
the frequencies of both insurance claims and power-grid disturbances are
higher in the winter months than in the summer months, but the
\rr{excess claim frequencies statistically associated with} geomagnetic activity follow similar trends
as for the full date range. The same is true when looking at the
subset of events attributed to surges in the low-voltage power
distribution grid.  

\rr{Figure~\ref{fig:e1} shows a similar diagram to that on lefthand side of
Fig.~\ref{fig:e}, now differentiating between the equinox periods and the solstice periods. Note that although the claims frequencies for the solstice periods are higher than those for the equinox periods, that difference is mainly a consequence of background (control) frequencies: the fractional excess frequencies on geomagnetically active days relative to the control dates are larger around the equinoxes than around the solstices.}

Fig.~\ref{fig:f} shows the comparison of claim ratios of
geomagnetically active dates relative to control dates for states with
high versus low geomagnetic latitude, revealing no significant contrast
\rr{(based on uncertainties computed as described above for Fig.~\ref{fig:g}).}

\section{Discussion and conclusions}\label{sec:conclusions}
We perform a statistical study of North-American insurance claims for
malfunctions of electronic and electrical equipment and for business
interruptions related to such malfunctions. We find that there is a
significant increase in claim frequencies in association with elevated
\ra{variability in the geomagnetic field}, comparable in magnitude to the increase in
occurrence frequencies of space weather-related disturbances in the
high-voltage power grid. In summary:
\begin{itemize}
\item The fraction of insurance claims \rr{statistically associated with} geomagnetic
  variability tends to increase with increasing activity from about
  $5-10$\%\ of claims for the top third of most active days to
  approximately 20\%\ for the most active few percent of days. 
\item The overall fraction of all insurance claims \rr{statistically associated with} the
  effects of geomagnetic activity is $\approx 4$\%. With a market
  share of about 8\%\ for Zurich NA in this area, we estimate
  that some 500 claims per year are involved overall in North America.
\item Disturbances in the high-voltage power grid \rr{statistically associated with}
  geomagnetic activity show a comparable frequency dependence on geomagnetic
  activity as do insurance claims. 
\item We find no significant dependence of the claims frequencies
  \rr{statistically associated with} geomagnetic activity on geomagnetic
  latitude.
\end{itemize}
\ra{For our study, we use a quantity that measures the rate of change of the geomagnetic
  field regardless of what drives that. Having established an impact
  of space weather on users of the electric power grid, a next step
  would be to see if it can be established what the relative
  importance of various drivers is (including variability in the ring current, electrojetc,
  substorm dynamics, solar insolation of the rotating Earth, \ldots),
  but that requires information on the times and
  locations of the impacts that is not available to us.}

The claims data available to us do not allow a direct estimate of the
financial impacts on industry of the malfunctioning equipment and the
business interruptions attributable to such malfunctions: \rr{we do
  not have access to the specific policy conditions from which each
  individual claim originated, so have no information on deductable
  amounts, whether (contingency) business interruptions were claimed
  or covered or were excluded from the policy, whether current value
  or replacement costs were covered, etc. Moreover, the full impact on
  society goes well beyond insured assets and business interruptions,
  of course, as business interruptions percolate through the complex
  of economic networks well outside of direct effects on the party
  submitting a claim.} A sound assessment of the economic impact of
space weather through the electrical power systems is a major
challenge, but we can make a rough order-of-magnitude estimate based
on existing other studies as follows.

The majority (59\%\ in sample B) of the insurance claims studied here are
explicitly attributed to ``Misc.: electrical surge'', which are
predominantly associated with quality or continuity of electrical
power in the low-voltage distribution networks to which the electrical
and electronic components are coupled. Many of the other stated causes
(see Section~\ref{sec:claimsdata}) may well be related to that, too,
but we cannot be certain given the brevity of the attributions and the
way in which these particular data are collected and recorded.  Knowing
that in most cases the damage on which the insurance claims are based
is attributable to perturbations in the low-voltage distribution
systems, however, suggests that we can look to a study that attempted
to quantify the economic impact of such perturbations on society.

That study, performed for the Consortium for Electric Infrastructure to
Support a Digital Society" (CEIDS) \citep{primen2001}, 
focused on the three sectors in the US economy that are
particularly influenced by electric power disturbances: the digital economy
(including telecommunications), the continuous process manufacturing
(including metals, chemicals, and paper), and the fabrication and
essential services sector (which includes transportation and water and
gas utilities). These three sectors contribute approximately 40\%\ of
the US Gross Domestic Product (GDP). 

\cite{primen2001} obtained information from a sampling of
985 out of a total of about 2 million businesses in these three
sectors. The surveys assessed impact by "direct costing" by combining
statistics on grid disturbances and estimates of costs of outage
scenarios via questionnaires completed by business
officials. Information was gathered on grid disturbances of any type
or duration, thus resulting in a rather complete assessment of the
economic impact. The resulting numbers were corrected for any later
actions to make up for lost productivity (actions with their own types
of benefits or costs).

For a typical year (excluding, for example, years with scheduled
rolling blackouts due to chronic shortages in electric power supply),
the total annual loss to outages in the sectors studied is estimated
to be {\$}46\,billion, and to power quality phenomena almost
{\$}7\,billion. Extrapolating from there to the impact on all
businesses in the US from all electric power disturbances results in
impacts ranging from {\$}119\,billion/year to {\$}188\,billion/year
(for about year-2000 economic conditions).

Combining the findings of that impact quantification of all problems
associated with electrical power with our present study on insurance
claims suggests that, for an average year, the economic impact of
power-quality variations related to elevated geomagnetic activity may
be a few percent of the total impact, or several billion dollars
annually. That very rough estimate obviously needs a rigorous
follow-up assessment, but its magnitude suggests that such a detailed,
multi-disciplinary study is well worth doing.


\begin{acknowledgments}
  We are grateful to the three reviewers of our manuscript for their
  guidance to improve the presentation of our results. This work was
  supported by Lockheed Martin Independent Research funds (CJS, SMP).
  The results presented in this paper rely on data collected at
  magnetic observatories. We thank the national institutes that
  support them and INTERMAGNET for promoting high standards of
  magnetic observatory practice (www.intermagnet.org). The insurance
  claims data are accessible at
  http://www.lmsal.com/$\sim$schryver/claims/claims.csv.
\end{acknowledgments}



%
%





\end{article}

\end{document}